\documentclass[manuscript]{aastex}

\slugcomment{To appear in Publications of the Astronomical Society
of the Pacific 2016}

\shorttitle{LOLAS-2: redesign of an optical turbulence profiler}
\shortauthors{Avila et al.}

\usepackage{paralist}
\usepackage{psfrag,color}
\usepackage{amsmath}
\usepackage[latin1]{inputenc}
\usepackage{natbib}
\usepackage{bm}
\usepackage{color}

\newcommand{\cn}{\ensuremath{C_{\mathrm N}^2}}
\newcommand{\cnh}{\ensuremath{C_{\mathrm N}^2(h)}}

\begin{document}

\title{LOLAS-2: redesign of an optical turbulence profiler with high altitude-resolution}

\author{
  R. Avila,\altaffilmark{1}
  C. A. Z\~uniga,\altaffilmark{1}
  J. J. Tapia-Rodr\'iguez,\altaffilmark{2,1}
  L. J. S\'anchez,\altaffilmark{3}
  I. Cruz-Gonz\'alez,\altaffilmark{3}
  J. L. Avil\'es,\altaffilmark{4}
  O. Vald\'es-Hern\'andez,\altaffilmark{1} and
  E. Carrasco\altaffilmark{4}}

\altaffiltext{1}{Centro de F\'isica Aplicada y Tecnolog\'ia Avanzada, UNAM, Campus Juriquilla, Apdo.
Postal 1-1010, 76000 Quer\'etaro, M\'exico.}

\altaffiltext{2}{Instituto Tecnol\'ogico de Morelia, Apdo. Postal 262, 58120 Morelia, Michoac\'an, M\'exico.}

\altaffiltext{3}{Instituto de Astronom\'ia, UNAM, Apdo. Postal 70-264, 04510 M\'exico D.F., M\'exico.}

\altaffiltext{4}{Instituto Nacional de Astrof\'isica, \'Optica y Electr\'onica, Apdo. Postal 51 y 216, 72840
San Andr\'es Cholula, Puebla, M\'exico.}

\begin{abstract}

We present the development, tests and first results of the second generation Low Layer Scidar (LOLAS-2).
This instrument constitutes a strongly improved version of the prototype Low Layer Scidar, which is aimed at the measurement of optical turbulence
profiles close to the ground, with high altitude-resolution.
The method is based on the Generalised Scidar principle which consists in taking double-star scintillation images on a defocused
pupil plane and calculating in real time the autocovariance of the scintillation.
The main components
are an open-truss 40-cm Ritchey-Chr\'etien telescope, a german-type equatorial mount, an Electron Multiplying CCD camera and a dedicated acquisition and real-time
data processing software. The new optical design of LOLAS-2 is significantly simplified compared with the prototype. The experiments carried out to test the permanence of the image within the useful zone of the detector and the stability of the telescope focus show that LOLAS-2 can function without the use of the autoguiding and autofocus algorithms that were developed for the prototype version. Optical turbulence profiles obtained with the new Low Layer Scidar have the best altitude-resolution ever achieved with Scidar-like techniques (6.3 m). The simplification of the optical layout and the improved mechanical properties of the telescope and mount make of LOLAS-2 a more  robust instrument.

\end{abstract}


\section{Introduction}
\label{sec:intro}

Turbulent flows in the atmosphere combined with stratified temperatures provoke turbulent fluctuations of the refractive index of air, commonly known as optical turbulence. The intensity of such fluctuations is determined by the second order structure constant {\cn}. The measurement of {\cnh},  $h$ being the altitude, is of major importance for the development of novel adaptive optical (AO) systems that overcome the angular-resolution degradation introduced by optical turbulence. The statistical study of optical turbulence profiles is also crucial for the characterization of sites where next generation of optical telescopes are to be installed (e.g., \citet{SER+09}).

One important limitation of AO systems that are designed to account for phase fluctuations generated all along the atmosphere is the tiny field of view over which the wavefront is corrected. One way to improve image quality over a wide field of view is to correct only the wavefront perturbations that come from turbulent layers close to the ground. This method is known as ground-layer adaptive optics (GLAO) \citep{Rig02,Tok04}. Indeed, the compensation of lower-altitude turbulent-layers provides wider corrected fields of view \citep{c98} and turbulence close to the ground is generally the most intense (e.g., \citet{AMV+04}). To develop a GLAO system for a given site, it is required to have as much knowledge as possible about the vertical distribution of optical turbulence in the ground layer (e.g, \citet{LH06}). This requires measurements of {\cnh} with very high altitude-resolution close to the ground.

Scintillation Detection and Ranging (SCIDAR) has extensively been used for {\cn} profiling since its invention by \citet{RRV74}. The generalised version of the SCIDAR made it possible to detect turbulence near the ground \citep{FTV98,AVM97}. A more recent implementation of the Generalized SCIDAR on a 40-cm telescope that used a widely-separated double star as a light source and an Electron Multiplying Charge Coupled Device (EMCCD) as the detector, gave place to the Low Layer SCIDAR (LOLAS) \citep{AAW+08}. The LOLAS prototype was used to characterize the ground-layer turbulence at Mauna Kea, together with a Slope Detection and Ranging instrument  \citep{CWA+09}. Although this several-years campaign gave definitive results, the experience showed that a number of aspects on the prototype LOLAS could be improved in order to optimize data acquisition.  In this paper we describe the development and tests of the second generation Low Layer Scidar (LOLAS-2).

A few instruments based on optical methods have been developed to measure {\cnh} profiles with the required vertical resolution close to the ground. For example, the surface-layer Slope Detection and Ranging (SL-SLODAR) \citep{OWB+10} makes use of two Shack-Hartmann wavefront sensors, each looking at one component of a widely-separated double star. The slope of the wavefronts coming from each star is measured on 5-cm square subapertures whereas LOLAS-2 measures scintillation on square elements of 1-cm side approximately, which makes the SL-SLODAR five times more sensitive than the LOLAS-2 for equal integration times or five times faster for equal number of photons per element. On the other hand, if the instruments were using the same double star separation, LOLAS-2 provides 5 times better altitude resolution. A similar concept that also uses a double star but measures the scintillation from each component on two cameras was implemented in the Stereo-Scidar by  \citet{SOW+14}. The Lunar Scintillometer, which consists of an array of photodiodes that measure scintillation from the moon, has been used to obtain high altitude-resolution turbulence profiles near the ground \citep{TBB10}. \citet{EM07} use a Generalized SCIDAR to distinguish layers at very similar altitudes but moving at different velocities.

The present paper is organized as follows: \S~\ref{sec:method} gives an overview of the LOLAS method. In \S~\ref{sec:lolasprototype} we briefly describe the prototype version of the instrument and in \S~\ref{sec:lolas2} the second generation of the instrument is presented. Tests on the instrument performances and some measurements are shown in \S~\ref{sec:instrumenttest} and  \S~\ref{sec:profiles}. Conclusions are given in \S~\ref{sec:conclusions}.

\section{Method}
\label{sec:method}

\begin{figure}[h!]
\epsscale{.50}
\plotone{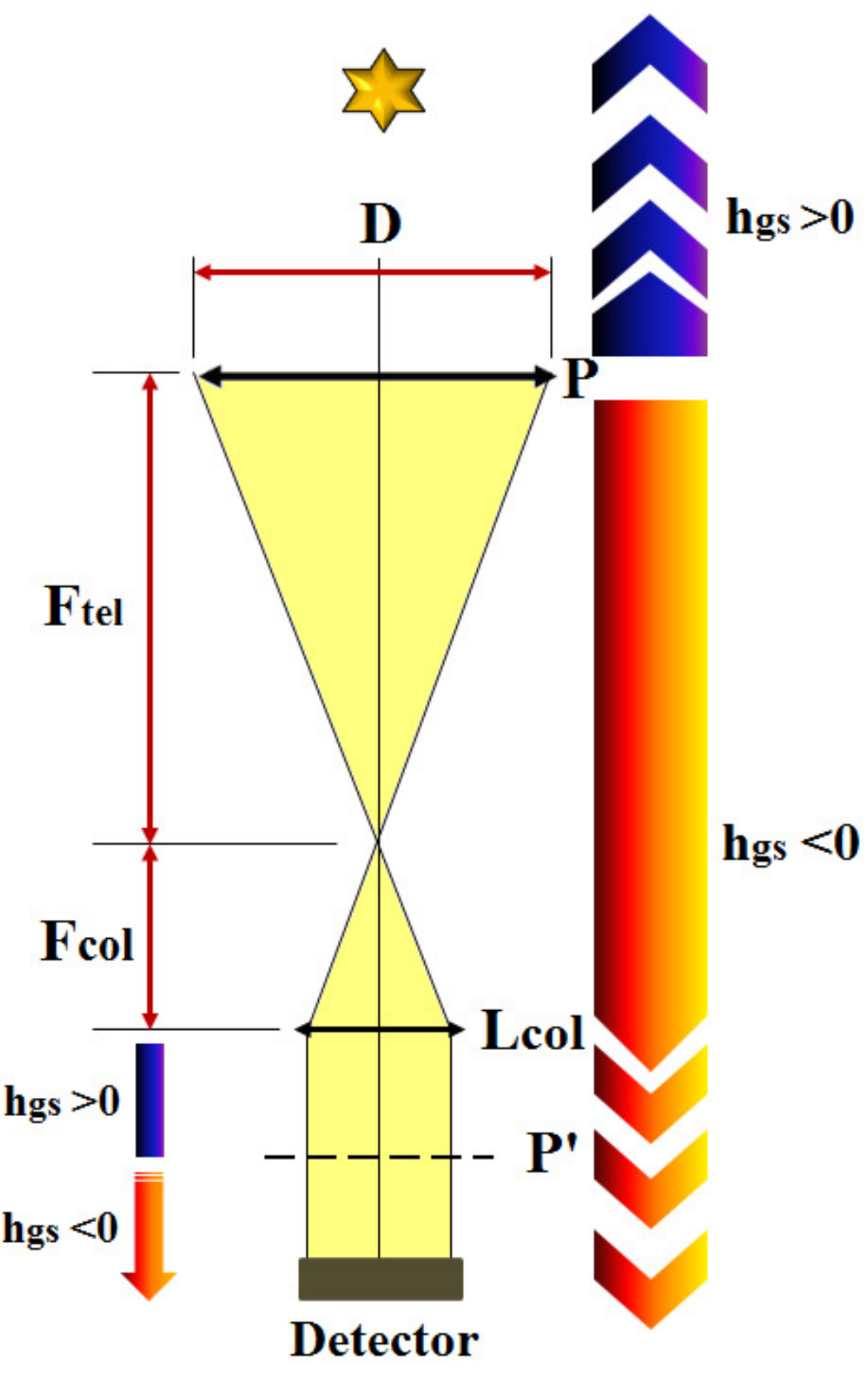}
\caption{Optical layout of the G-SCIDAR.}
\label{fig:SCIDAR1}
\end{figure}

The Low Layer Scidar is based on the principle of the Generalized
Scintillation Detection and Ranging (G-SCIDAR) technique
\citep{RRV74,FTV98,AVM97}. In this section, we present a brief overview of
the LOLAS method, since a complete description can be found in
\citet{AAW+08}. The G-SCIDAR principle can be summarized as
follows: double-star scintillation patterns are recorded on short
exposure-time images on a virtual plane located a distance
$h_\mathrm{gs}$ from the telescope pupil. In G-SCIDAR experiments
this virtual plane is located below the telescope pupil, thus
$h_\mathrm{gs}<0$. A schematic view of the optical setup is shown
in Fig. \ref{fig:SCIDAR1}. Each image consists of a randomly
distributed intensity pattern. The autocovariance of this
stochastic illumination is obtained by computing the spatial
normalized autocorrelation of each image and averaging those
autocorrelations over thousands of statistically independent image
samples. Each layer of optical turbulence contributes to the
resulting scintillation autocovariance with three covariance
peaks: one centred at the autocovariance origin and  two others,
identical to each other,  separated from the origin by
$\mathbf{d}_\mathrm{l}=-\boldsymbol{\rho} \lvert h-h_\mathrm{gs}
\rvert$ and $\mathbf{d}_\mathrm{r}=\boldsymbol{\rho} \lvert
h-h_\mathrm{gs} \rvert$, respectively, where $\boldsymbol{\rho}$
denotes the angular separation of the double star and $h$ the
layer altitude above the ground. Knowing $\boldsymbol{\rho}$ and
the conjugation altitude $h_\mathrm{gs}$, the experimental
determination of $\mathbf{d}_\mathrm{l}$ (or
$\mathbf{d}_\mathrm{r}$) leads to an estimate of the layer
altitude $h$. The measured autocovariance peaks
$C_\mathrm{gs}(\mathrm{\mathbf{r}})$ are proportional  to the optical
turbulence strength at altitude $h$, $C_\mathrm{N}^2(h)$, and to
the scintillation autocovariance function $K(\mathbf{r}, \lvert
h-h_\mathrm{gs} \rvert)$. In the realistic case of multiple
layers, the response of each layer adds up to the measured
autocovariance, resulting in Eq.~1 of \citet{AAW+08}. This equation consists of an
integral with respect the altitude $h$ that is similar to a
convolution, except that the kernel  $K(\mathbf{r}, \lvert
h-h_\mathrm{gs} \rvert)$ depends on the integration variable. One
needs to invert this integral equation to retrieve
$C_\mathrm{N}^2(h)$. Due to its similarity with a convolution
integral, we developed an algorithm based on the CLEAN method, but
in which the kernel is recalculated for each different value of
$h$. Our modified CLEAN algorithm is based on that reported by
\citet{PDA01}.  As explained by \citet{AAW+08}, the achievable
altitude resolution using the modified CLEAN method, when the target is at the zenith is $\Delta
h=0.52\sqrt{\lambda(\lvert h-h_\mathrm{gs} \rvert)}/\rho$, where
$\lambda$ is the wavelength and $\rho=\lvert \boldsymbol{\rho}\rvert$. When the star is located at an elevation angle $\theta$, $\Delta h$ is decreased by a factor $\cos \theta$.
We take $\lambda=0.5 \mu\mathrm{m}$,
which corresponds to the maximum sensitivity of our detector.  For
a telescope having an aperture $D$, the maximum altitude for which
the $\cn$ value can be measured is given by $h_\mathrm{max}=D/\rho$
\citep{AAW+08}.

\bigskip

The Low Layer Scidar concept consists of putting into practice a
G-SCIDAR on a dedicated portable telescope, using widely separated
double stars as light sources. As can be seen from the above
expressions for $\Delta h$ and $h_\mathrm{max}$, the wider the
separation, the better the altitude resolution but the shorter the
maximum altitude. LOLAS was designed to use a 40-cm telescope, an
EMCCD to improve
sensitivity and a real-time computation of the scintillation
autocovariance. Sections \ref{sec:lolasprototype} and
\ref{sec:lolas2} describe the prototype LOLAS version and the
second generation instrument, respectively.

\section{LOLAS Instrument development}
\label{sec:lolasinst}

\subsection{Prototype version}
\label{sec:lolasprototype}

\begin{figure}[h!]
\epsscale{.50}
\plotone{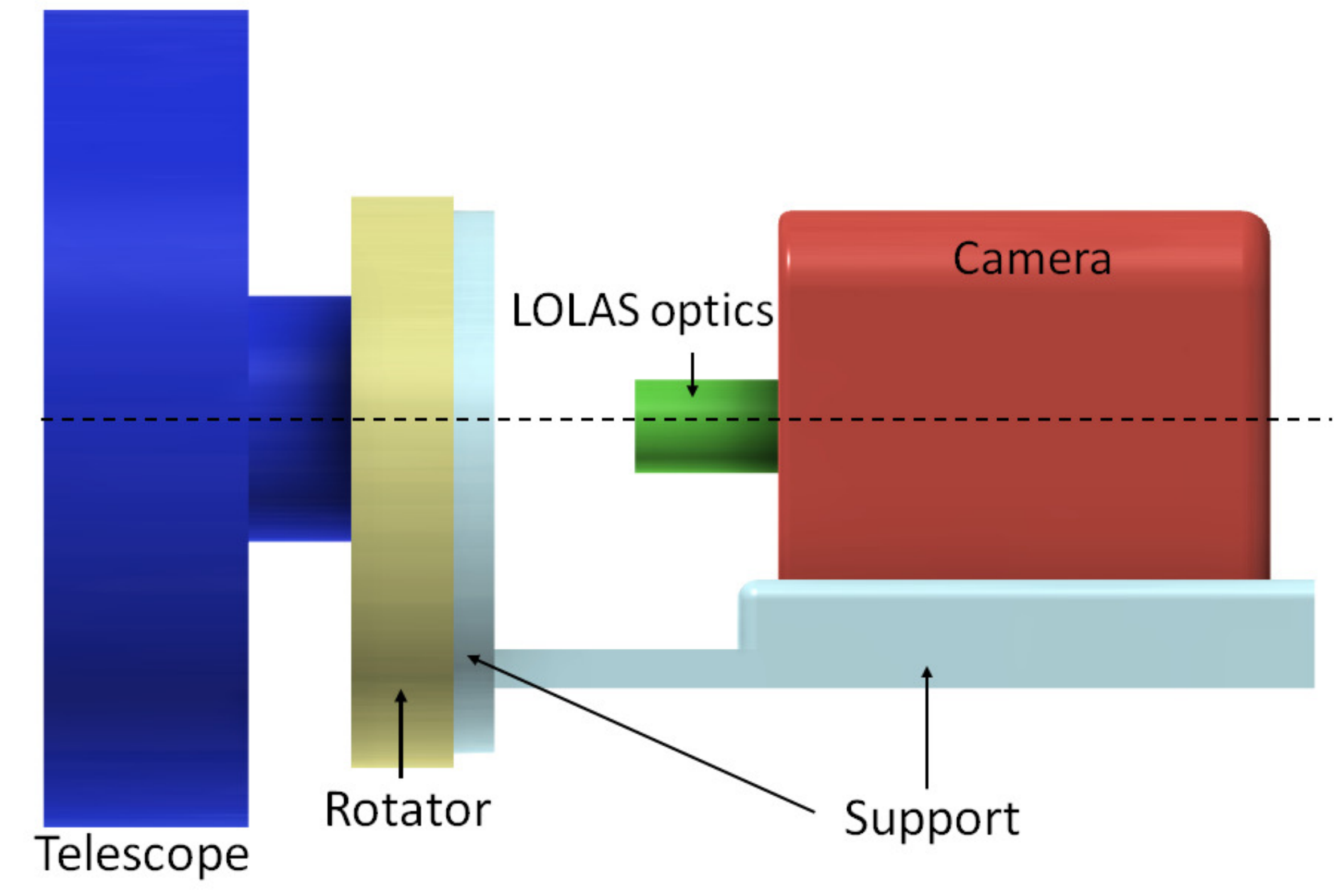}
\caption{Schematic view of the prototype LOLAS
instrument.} \label{fig:ProLOLAS}
\end{figure}

The instrumental setup of LOLAS has been widely explained elsewhere
\citep{AAW+08,CWA+09,AAB+12}. We present a summary of the most important instrumental
characteristics. Figure \ref{fig:ProLOLAS} shows a schematic view of the
prototype LOLAS. The scintillation images are obtained with a
Schmidt-Cassegrain telescope of focal ratio f/10 and
diameter $D = 40.64$~cm and installed on an equatorial mount,
manufactured by Meade. The optics consists of two achromatic
lenses of 50~mm focal-length.  With this optical arrangement, the virtual analysis plane
is located  1.94~km  below the pupil. The diameter of the pupil
image on the detector is $D^\prime = 24.5$~mm. The scintillation
images are captured by an EMCCD camera (Andor iXon) with
$512\times 512$ square pixels of $16~\mu\mathrm{m}$.
The frames are binned 2 $\times$ 2, and the active zone is limited
to an array of $256 \times 80$ binned pixels. The exposure time of
each frame ranges from 3 to 10 ms, depending on the wind
conditions. The typical number of images to obtain one
autocovariance is set to 30000.

The EMCCD camera  is mounted on a base that is attached to the
rear of the telescope (see Fig. \ref{fig:ProLOLAS}). The same
equipment, but with different optics, is used to form the SLODAR
instrument. To switch between each instrument with the required
positioning accuracy, a manual exchange mechanism was installed in
front of the camera.

\subsection{Second generation}
\label{sec:lolas2}

\begin{figure}[h]
\epsscale{.50}
\plotone{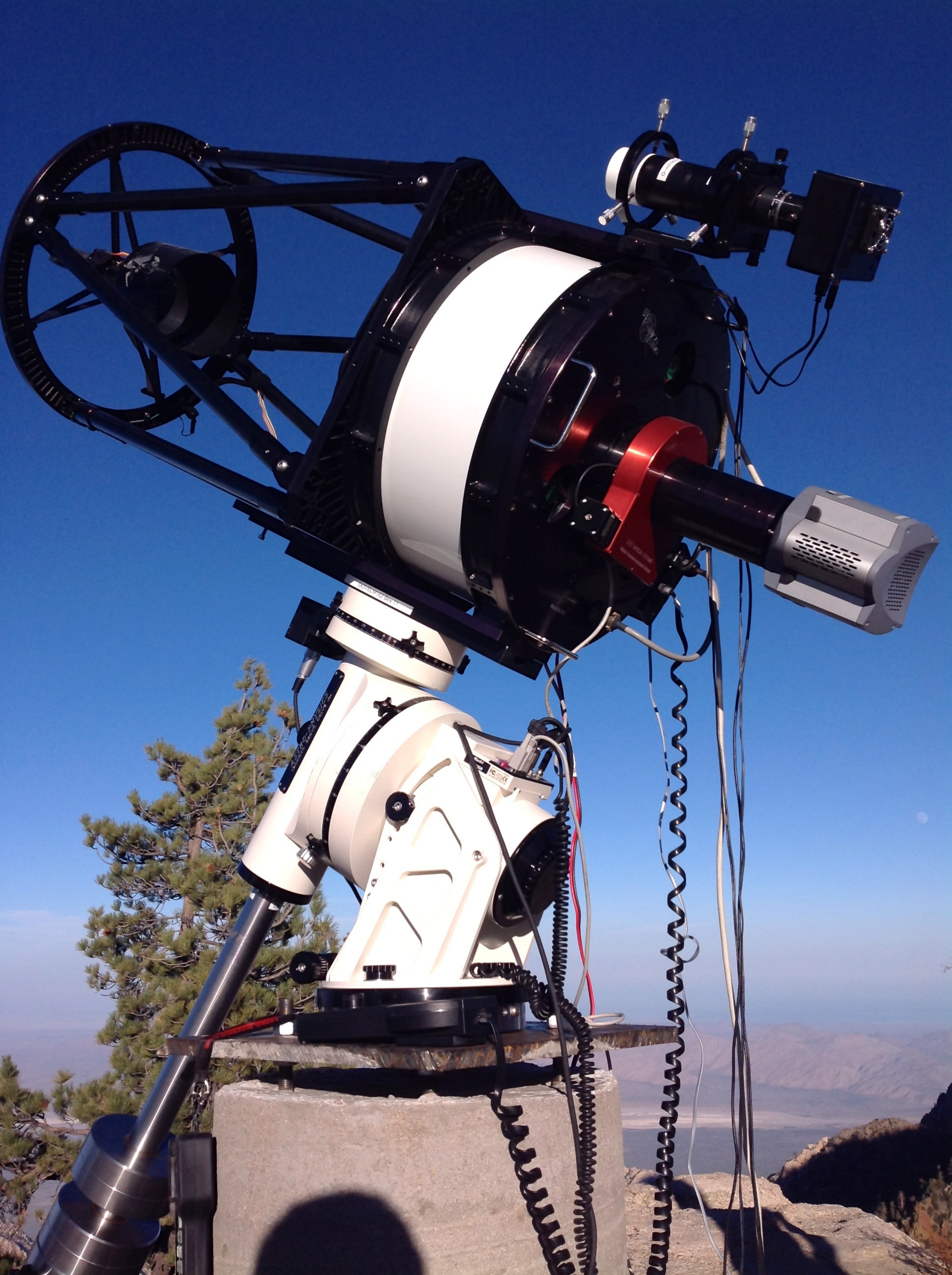}
\caption{LOLAS-2 at the Observatorio Astron\'omico Nacional de San Pedro M\'artir (OAN-SPM).} \label{fig:InstLOLAS}
\end{figure}

As seen in Fig. \ref{fig:InstLOLAS}, LOLAS-2 uses a Ritchey-Chr\'etien open telescope of focal ratio
f/9 and diameter $D = 40.64$~cm, manufactured by RC Optical
Systems, a German equatorial mount (1200GTO) manufactured by
Astro-Physics, an EMCCD camera (Andor iXon) to acquire scintillation images and a Sbig St-402ME
camera for the finder telescope.

One advantage of using a Ritchey-Chr\'etien telescope is that when
the position of the focal plane is changed, the effective focal
length of the telescope remains unchanged. Our RC Optical Systems
telescope is equipped with a system that maintains the focus by
monitoring the secondary mirror position in closed loop at a
frequency of 6 kHz, reaching an accuracy in the secondary mirror
and the focus positions of 0.6 and 25.4 $\mu\mathrm{m}$,
respectively. This control system is part of the telescope. In
addition, the fact that the telescope is open prevents air at
different temperatures to get trapped inside the tube in a
turbulent convective flow, which would add an instrumental bias to
the  $\cn$ measurements at ground level. This was the case with
the prototype LOLAS. Removal of the spurious  turbulence from the
measurements was performed in a post-processing procedure using
the method described by \citet{AVS01}, as reported by
\citet{CWA+09}. In LOLAS-2, this post-processing step is avoided,
making the data reduction faster and simpler.

Concerning the mount, LOLAS-2 incorporates a German-type mount
that reduces considerably the lever arm between the equatorial and
declination axis. Moreover, the Astro-Physics mount has a higher
stiffness and the worm gear accuracy is significantly better,
compared to those of the Meade mount.

\begin{figure*}[h!]
\epsscale{1.1}
\newlength\thisfigwidth
\setlength\thisfigwidth{0.5\linewidth}
\addtolength\thisfigwidth{-2 cm}
\hfill
\makebox[\thisfigwidth][l]{{\hspace{2cm}\textbf{a}}}%
\hfill%
\makebox[\thisfigwidth][l]{{\hspace{2cm}\textbf{b}}}\\[3ex]
\plottwo{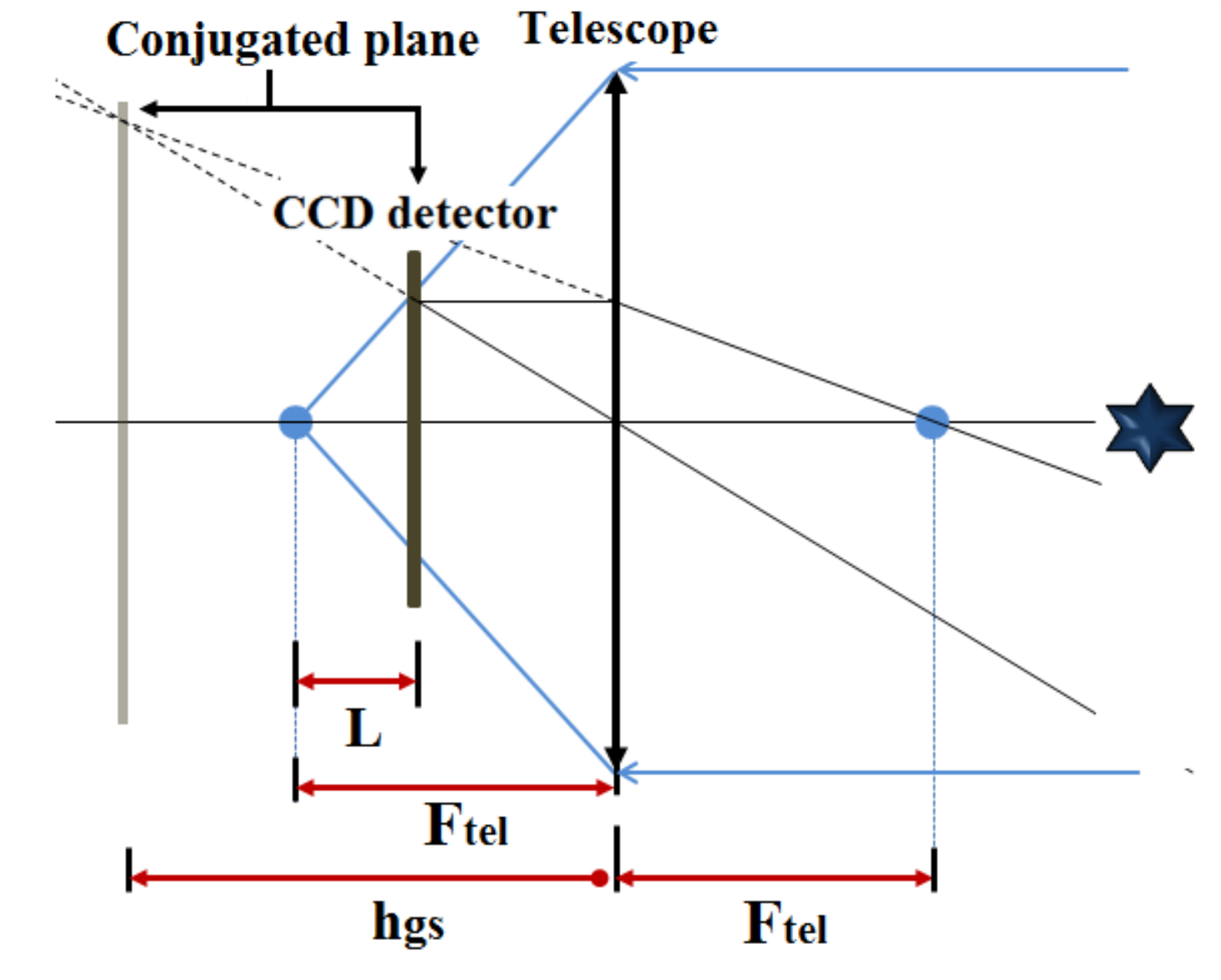}{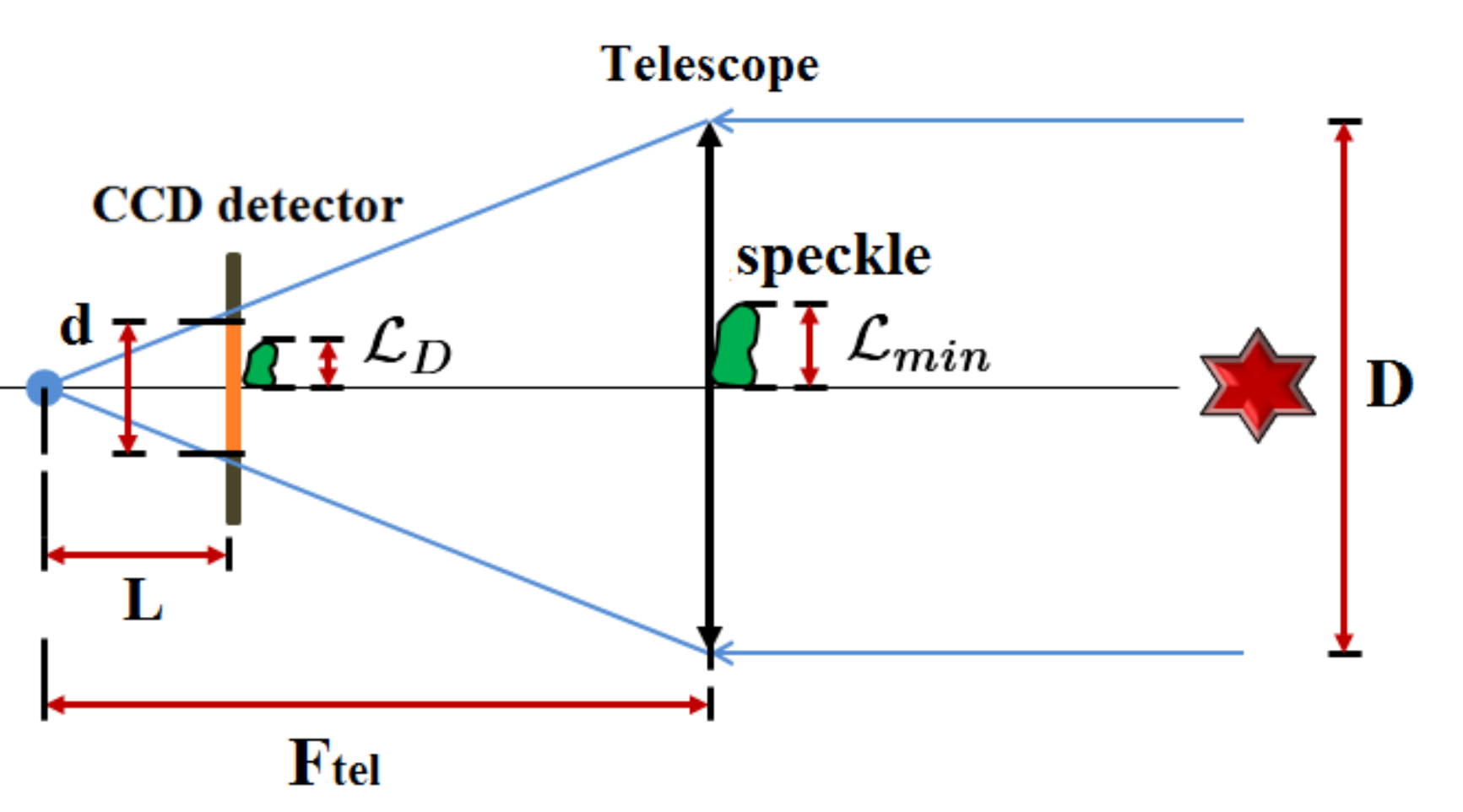}
\caption{(\textit{a}) Optical scheme of LOLAS-2. When the
CCD is placed before the focal point of the telescope, its conjugate plane is located a distance $h_ \mathrm{gs}$ underneath the pupil ($h_ \mathrm{gs} < 0$), (\textit{b}) Scheme showing the geometrical relations that lead to Eq.~\ref{Ec:ConservSpeck}.}
\label{fig:widefig2}
\end{figure*}

The prototype version uses the optics of the telescope and
achromatic doublets to define the spatial sampling and conjugation
distance $h_{gs}$ below the pupil. The second generation
instrument was developed so as to dispense the use of the
achromatic doublets and the exchange mechanism. It only uses the
telescope optics, making it a simpler and more robust instrument.
The beam is no longer collimated, like in the prototype version.
The EMCDD is placed directly a distance $L$ before the telescope
focal plane. Distance $L$ is chosen such that the detector plane
is made the conjugate of the a virtual plane located a distance
$h_\mathrm{gs}$  below the telescope pupil (see Fig.
\ref{fig:widefig2}{\it a}) and the spatial sampling on this plane is
well-suited to sample the scintillation speckles (see Fig.
\ref{fig:widefig2}{\it b}).  Using the thin lens equation, it can easily
be shown that $h_\mathrm{gs}$  is related to $L$ by the following
expression:
\begin{equation}
h_\mathrm{gs}  = -\frac{{F_\mathrm{tel}}^2-F_\mathrm{tel}L}{L},
\label{Ec:ThinLen}
\end{equation}
where $F_\mathrm{tel}$ is the focal length of the telescope. For
the spatial sampling, Fig. \ref{fig:widefig2}{\it b} illustrates the
demagnification relation:
\begin{equation}
\frac{\mathcal{L}_{D}}{L}= \frac{\mathcal{L}_\mathrm{min}}{F_\mathrm{tel}}
\label{Ec:ConservSpeck}
\end{equation}
where $\mathcal{L}_\mathrm{min}$ represents the typical size of
the smallest scintillation speckles on the pupil and
$\mathcal{L}_{D}$ is its corresponding size on the detector plane.
\citet{PDA01} showed that the full width at half maximum of the
autocovariance of the scintillation produced at altitude $h$ is
given by
\begin{equation}
\mathcal{L}(h)=0.78\sqrt{\lambda \lvert h-h_\mathrm{gs} \rvert)}.
\label{Ec:FWHM}
\end{equation}
This is, the typical size of the smallest speckle is
$\mathcal{L}_\mathrm{min}\equiv \mathcal{L}(0)=0.78\sqrt{\lambda
\lvert h_\mathrm{gs}\rvert}$. Solving Eq.~\ref{Ec:ConservSpeck}
for $\mathcal{L}_{D}$ and replacing the above expression for
$\mathcal{L}_\mathrm{min}$ gives:
\begin{equation}
\mathcal{L}_{D}= L\frac{0.78\sqrt{\lambda  \lvert h_\mathrm{gs}\rvert}}{F_\mathrm{tel}}.
\label{Ec:SpatSamp}
\end{equation}
For ground-level turbulence to be detectable,
$h_\mathrm{gs}$ must be smaller than $-1000$~ m. A good spatial sampling of the
scintillation speckles is obtained when $\mathcal{L}_{D}\simeq
2p$, where $p$ is the size of  the elementary sampling element. We
chose to acquire images with the camera pixels binned $2\times2$.
In that case, $p=2d_\mathrm{pix}$. The camera pixel size is
$d_\mathrm{pix}=16\;\mu\mathrm{m}$. Table  \ref{tab:ComparisonL}
gives values of different parameters of interest for different
values of $L$. It can be seen that a good compromise is obtained
when the detector is located a distance $L=11$~ mm before the
telescope focal plane, as the conjugation distance is large enough
($h_\mathrm{gs} = -1212$~ m) and the number of spatial samples per
smallest speckle width is 1.8, which is very close to 2, the Nyquist
criterion. Even though the undersampling is small, it is taken into account in the kernel of the inversion process that calculates {\cn} profiles from the measured autocovariances.

As a consequence of this analysis, the value for $L$ is set to 11 mm in LOLAS-2.
For this value of $L$, the distance separating two contiguous binned pixels corresponds to an angular separation of $1.8^{\prime\prime}$.

\section{Instrument performance}
\label{sec:instrumenttest}

In this section we present results concerning the focus stability
and guiding performance obtained the Observatorio Astron\'omico
Nacional at San Pedro M\'artir (OAN-SPM), Baja California, Mexico. The data
was obtained on the nights of 2013 June 15, 16 and 17. Table
\ref{tab:UsedStars} summarises the pertinent characteristics of
the double-star targets used.

\subsection{Focus stability}
\label{sec:focus}

\begin{figure} [h!]
\epsscale{.70}
\plotone{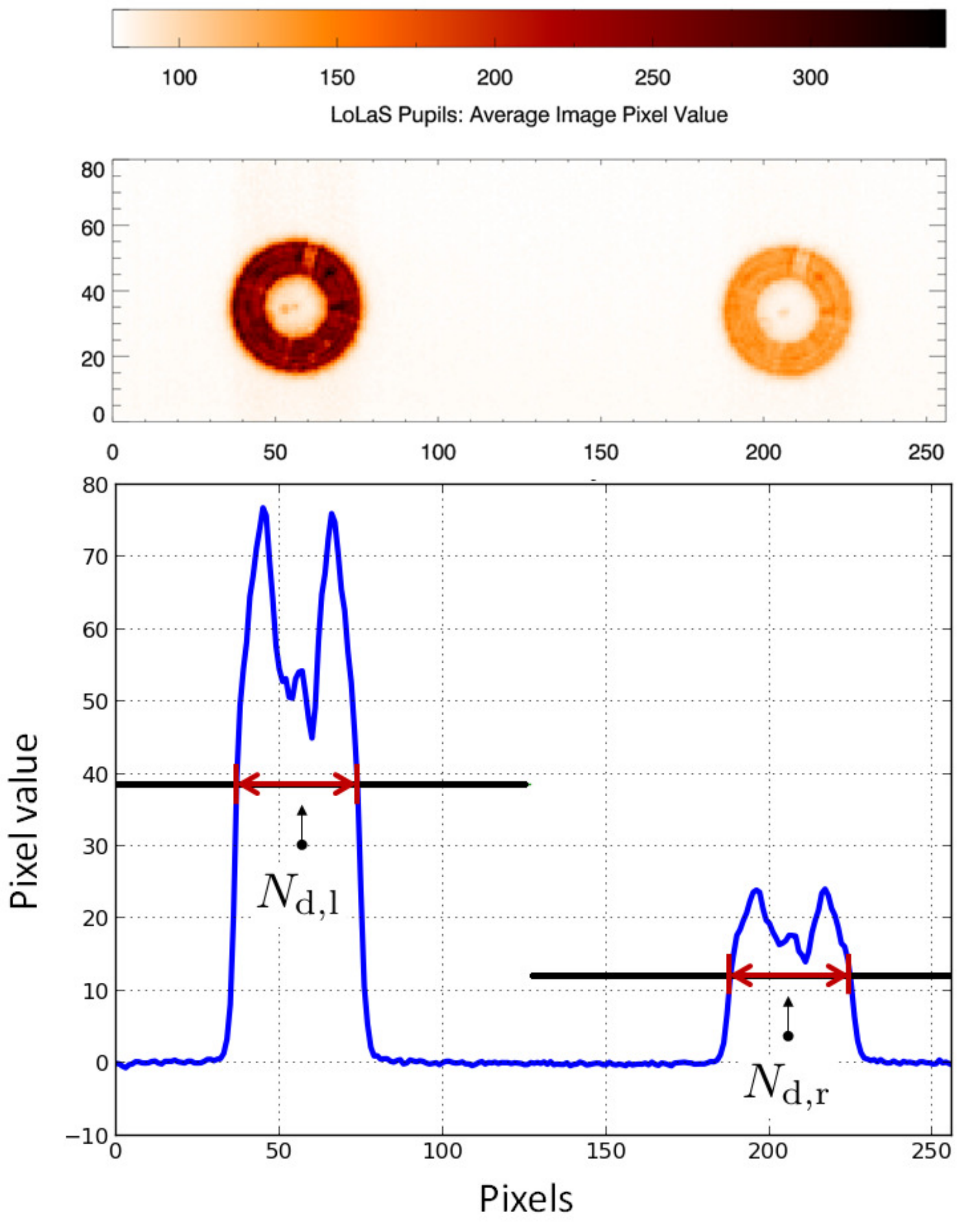}
\caption{Average image for the calculation of pupil diameters $N_{d,l}$ and $N_{d,r}$ in pixels. }
\label{fig:ImageMean}
\end{figure}

Focus stability is extremely important to maintain the spatial
sampling and conjugation altitude along data acquisition. A
variation of the telescope focus position translates into a
variation of the pupil image diameter $d$ on the detector. This
diameter is continuously monitored during the standard data
acquisition of the instrument. Images are sent by the EMCCD to the
computer in packets of 200 consecutive 256x80 pixels frames. In
each frame, the image is centred (as explained in \S
\ref{sec:guiding}) and then co-added to form a mean image made of
200 frames. The pupil diameters are calculated from this mean
image as follows (see Fig. \ref{fig:ImageMean}): the mean image is
integrated along columns to form a row of the accumulated values.
The width at half maximum of the left and right pupils on the
accumulated-values row determines their diameter in number of
pixels $N_{d,l}$ and $N_{d,r}$, respectively.

To estimate the focus stability, the diameter of each pupil was monitored
during several hours on three nights, using the same double star as source.
The total number of pupil size measurements was 4566.
The mean and standard deviation of the measured diameters are $\left<N_{d}\right>=38.13$ and $\sigma_{N_{d}}=0.40$ pixels.  The average of the temperature and its variation for each night was
 $13.5\pm0.22$, $14.6\pm0.23$ and $14.1\pm0.07$ Celcius degrees, according to the weather station at SPM.
The obtained
standard deviations can be due to uncertainties in the estimation
procedure and/or to actual physical variation in the focus or
camera positions. If we consider the latter to be the cause of the measured diameter fluctuations, the
consequence of the deviation of 0.40 pixels in pupil diameter
implies a variation of 13 m in altitude conjugation and 0.32
$\mu$m in the sampling element size on the detector,
approximately, which are tolerable values. This result suggests
that the use of an auto-focus algorithm, which was developed for
the prototype LOLAS,  could be avoided in the second generation
instrument, although more tests under varying temperature conditions should be performed.

\subsection{Guiding test}
\label{sec:guiding}

Maintaining a constant position of the pupil images on the EMCCD
is important to correctly compute the mean image, the
autocorrelation of which is used to normalize the mean
autocorrelation of scintillation images, so obtaining the
scintillation autocovariance. On wind conditions commonly
encountered in astronomical observatories, telescope shake may
cause pupil images to move on the detector. This eventual image
wander is corrected for by centering the image in every single
frame captured by the detector. The guiding performance of the
telescope mount is important to keep pupil images within the
useful window of the EMCCD during data acquisition that may last
hours.

The image position on the active area of the EMCCD is determined
as follows: we construct an artificial reference image
$I_\mathrm{r}(\mathbf{r})$ formed by two disks of 38 pixels in
diameter each and separated from each other by the same distance
as the observed double star. We compute the cross-correlation
$C_\mathrm{c}(\mathbf{r})$ of the current image $I_i(\mathbf{r})$
with $I_\mathrm{r}(\mathbf{r})$. The position of the pupil images
$(X,Y)$ in  $I_i(\mathbf{r})$ is set as the position of the
maximum value of $C_\mathrm{c}(\mathbf{r})$ with respect the
frame centre.

\begin{figure}[h!]
\epsscale{.70}
\plotone{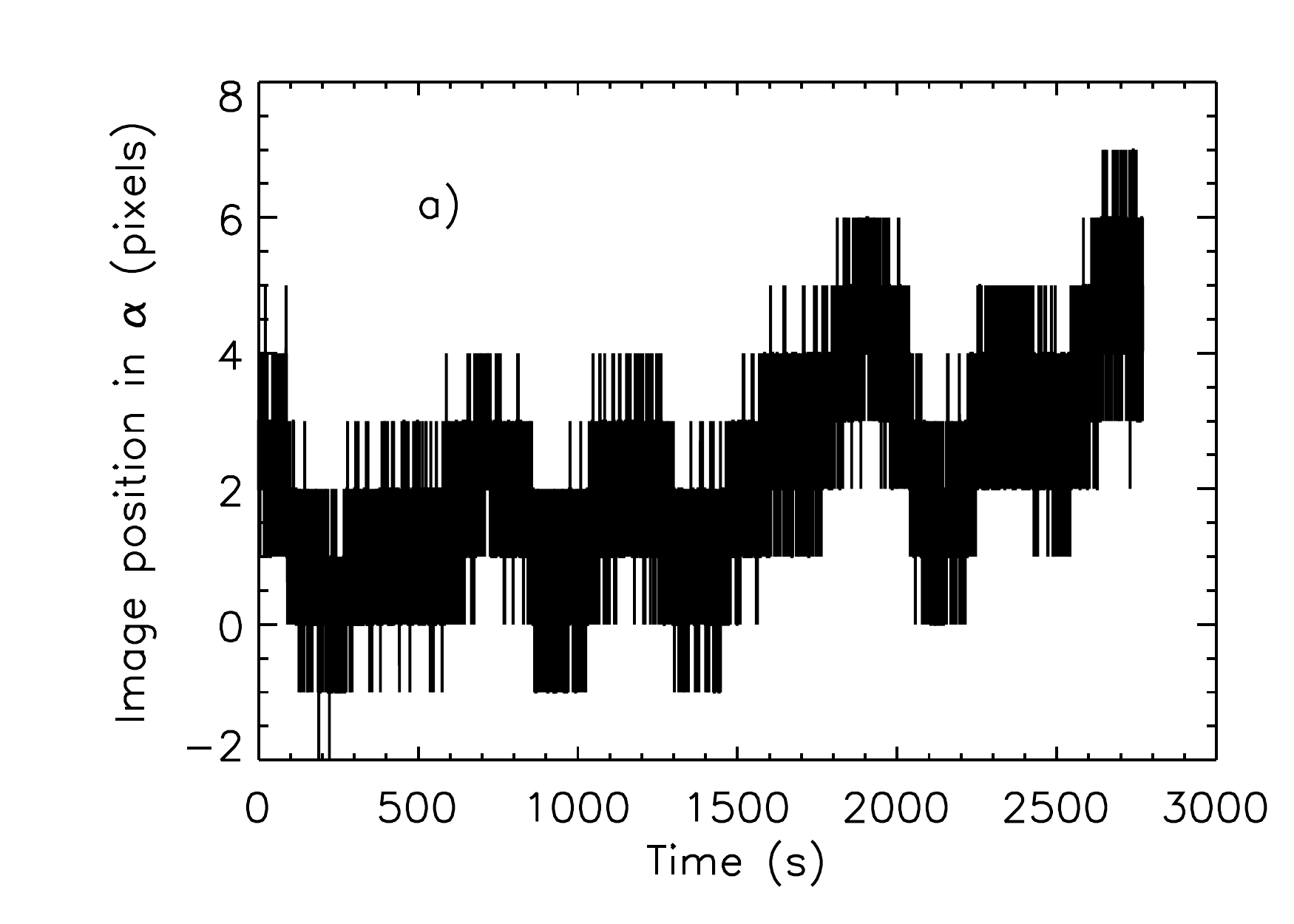}
\plotone{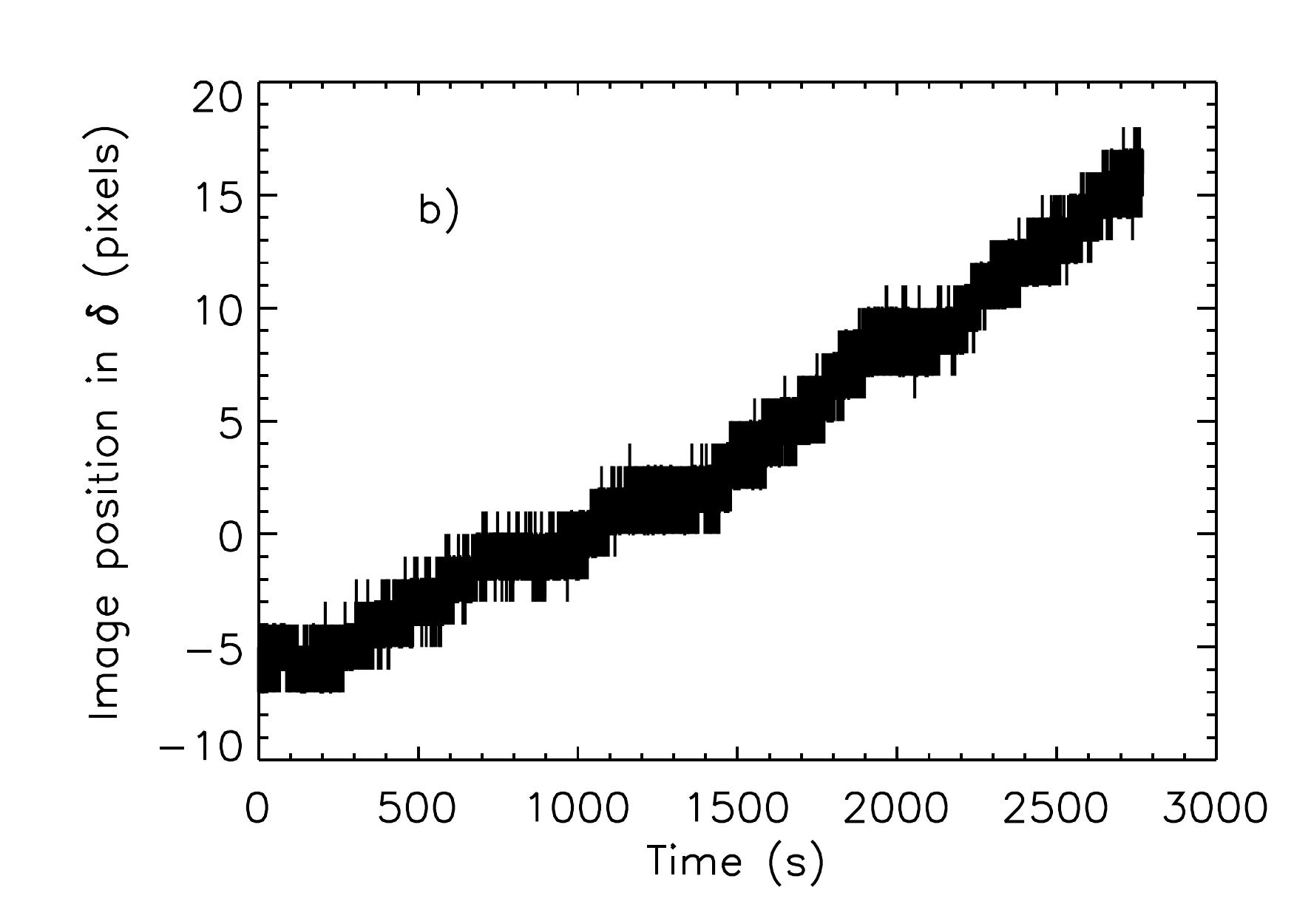}
\caption{Example of recorded image positions in pixels with respect to the
EMCCD centre as a function of time, along right ascension (\textit{a}) and
declination (\textit{b}) directions, respectively. Data were obtained on 2014 June 15.} \label{fig:Offset2013-1}
\end{figure}

\begin{figure}[h!]
\epsscale{.70}
\plotone{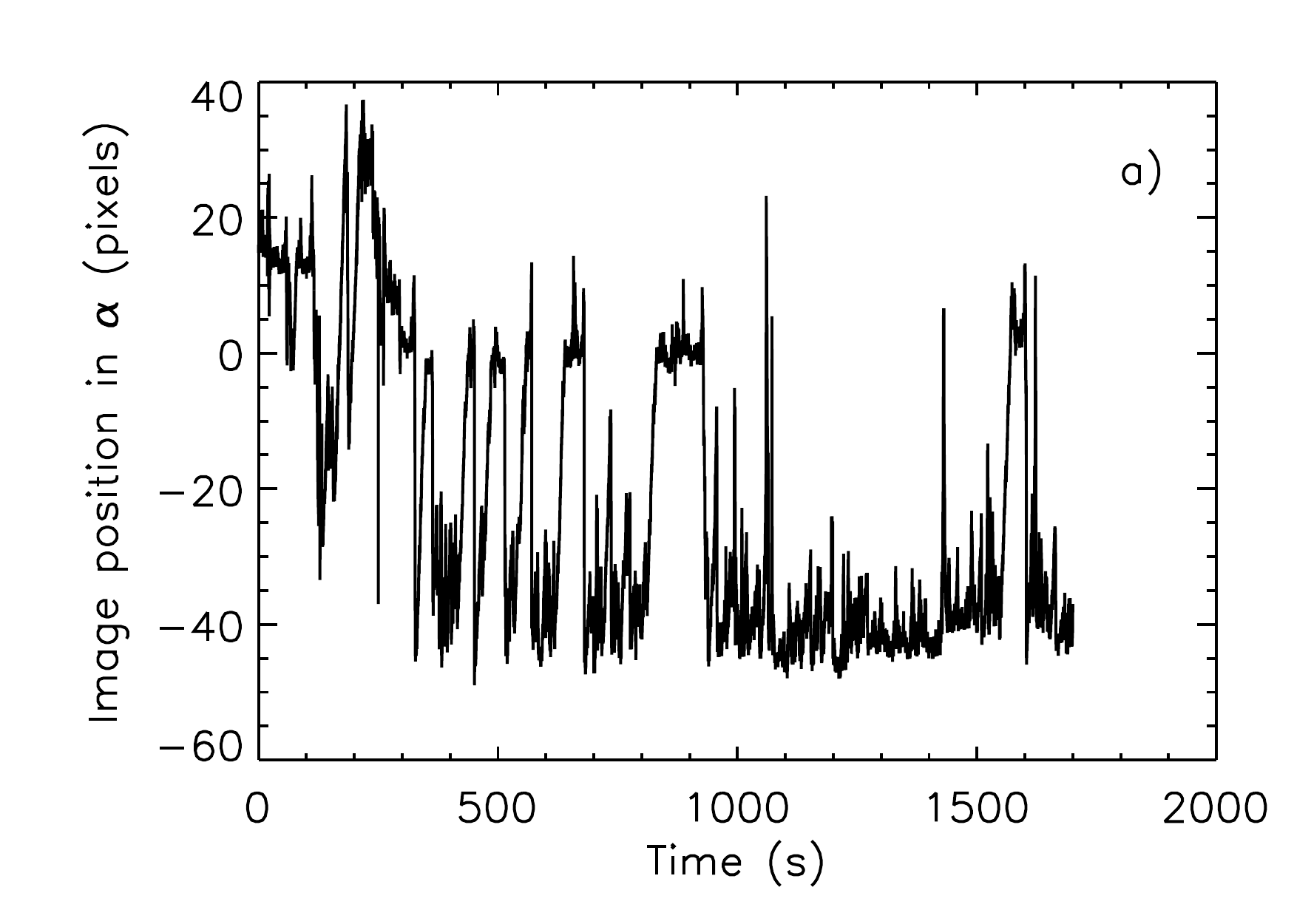}
\plotone{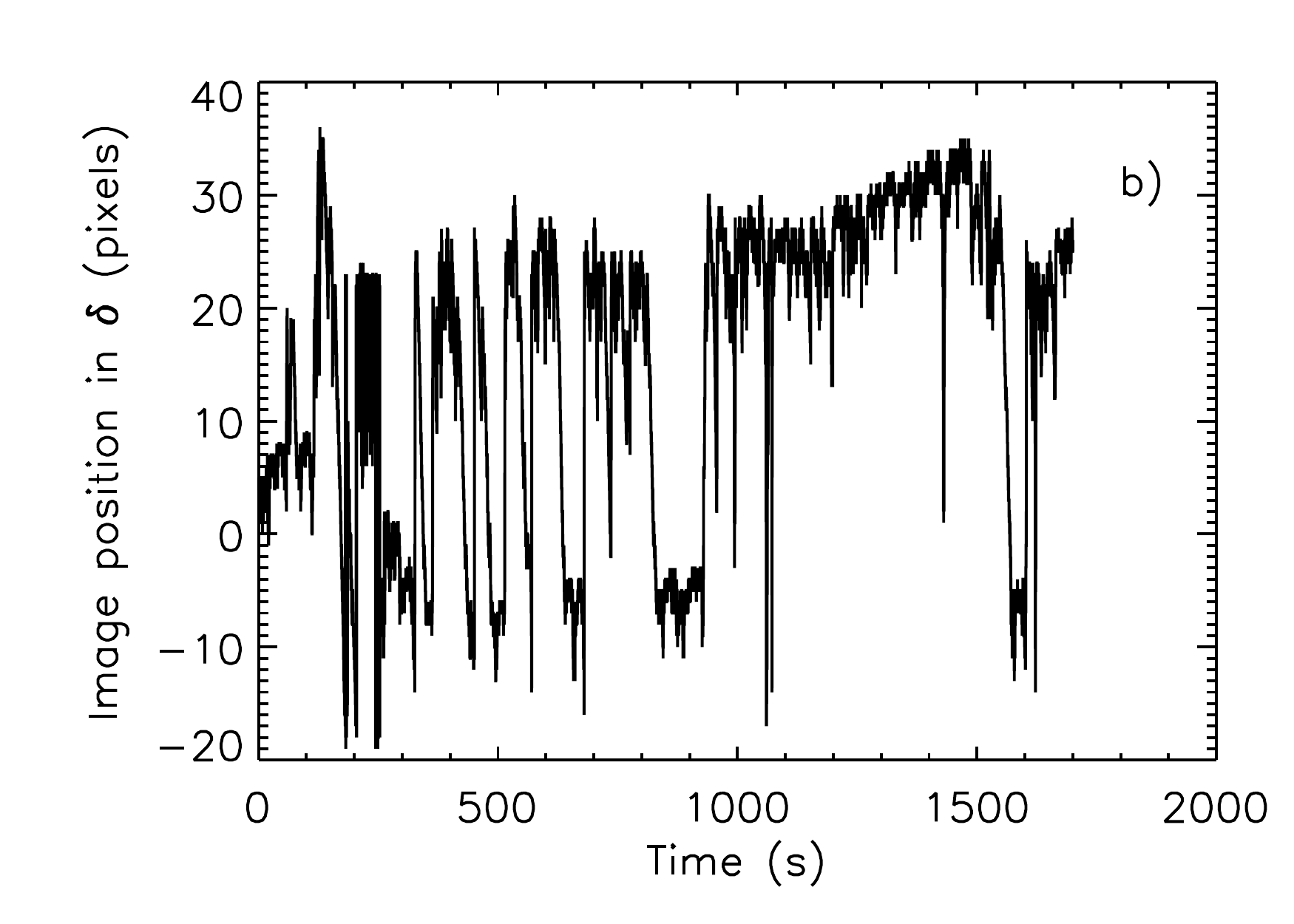}
\caption{Similar to Fig.~\ref{fig:Offset2013-1} for data obtained on 2014 June 17.} \label{fig:Offset2013-2}
\end{figure}

In standard operation of LOLAS-2, image positions are not saved on
disk. To test the guiding performance and image stability we recorded image positions
$(X,Y)$ during several observations. The $(X,Y)$ coordinates on the EMCCD were
rotated according to the double-star position angle to obtain image positions in
right ascension ($\alpha$) and declination ($\delta$) coordinates system. This allows us
to investigate separately the guiding and stability behaviour on the two rotation axis of the mount.
Figures~\ref{fig:Offset2013-1} and~\ref{fig:Offset2013-2}
show two examples. For the results obtained on 2013 June 17 UT (Fig.~\ref{fig:Offset2013-1}) wind was blowing from the South-West
with mean speed of 15.5 km h$^{-1}$ and a maximum value of 18.0 km h$^{-1}$.
The image positions remained very stable apart from a slow continuous drift on $\delta$ and
a slow oscillation in $\alpha$, presumably due to an inexact alignment of the mount and a periodic error on the right-ascension gear-mechanism, respectively.
On 2013 June 15 UT wind was less benevolent, blowing from the West at a mean speed of 26.6  km h$^{-1}$ with gusts reaching  40 km h$^{-1}$. Figure~\ref{fig:Offset2013-2} shows an
example of that night. Even though images were moving
significantly, the mean position remained constant and pupil
images remained within the EMCCD working window, which enabled
turbulence profiles to be measured in these conditions.

\begin{figure}[h!]
\epsscale{.70}
\plotone{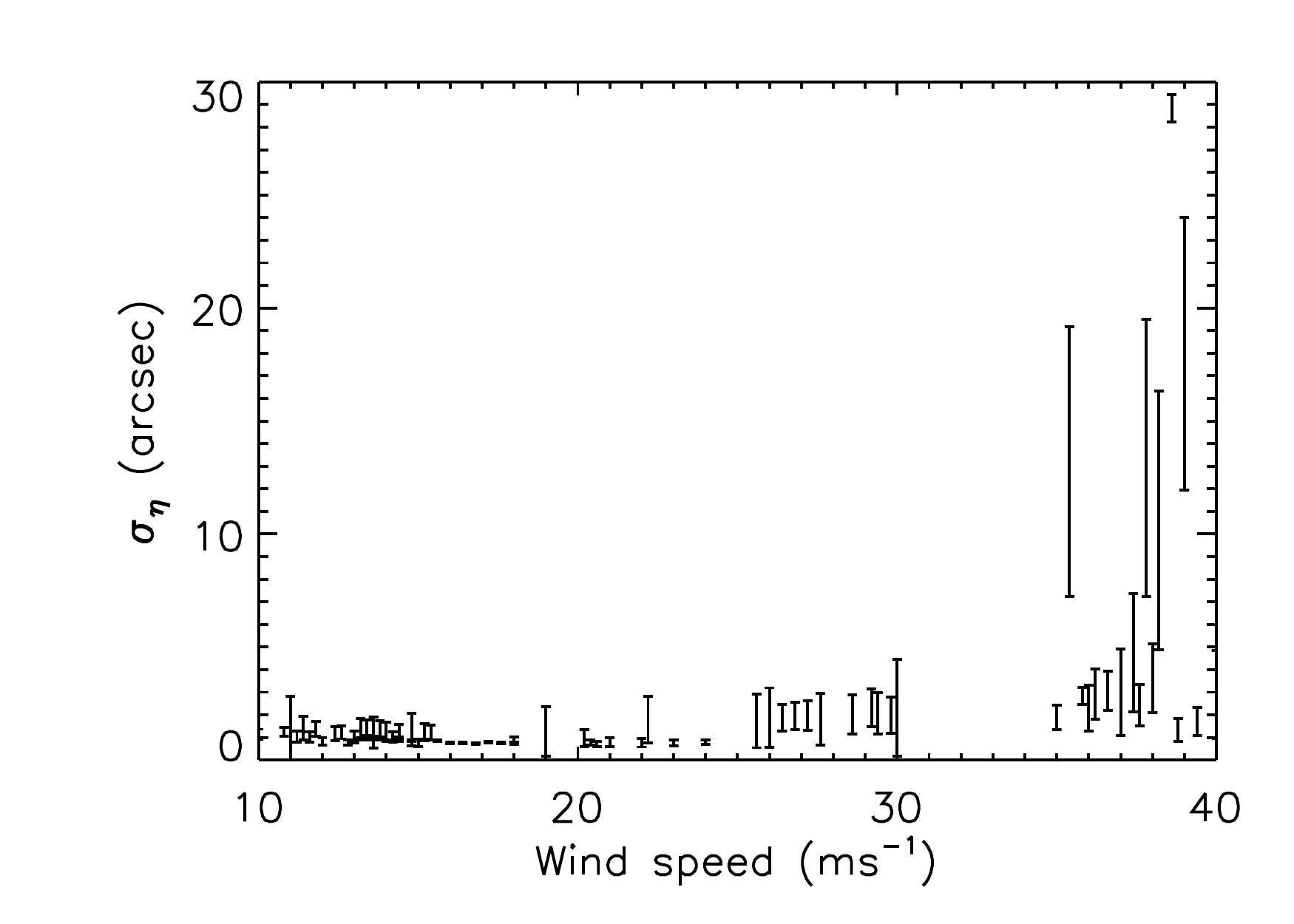}
\caption{Image jitter as a function of wind speed. Bars indicate the 25 and 75 percentiles of the image jitter (standard deviation of image position) values measured under a given value of wind speed.}
\label{fig:jitter-wind}
\end{figure}

To investigate further the dependence of image jitter on prevailing wind speed, in Fig.~\ref{fig:jitter-wind} it  is shown a plot of the standard deviation of image position as a function of wind speed. For each of the available 4566 frame-packets, the standard deviation $\sigma_\eta$ of the image position was calculated. The frame rate within each packet is 13-ms per frame. The wind conditions that prevailed at the time of each measurement were obtained from the database of the OAN-SPM weather station\footnote{www.astrossp.unam.mx/weather15/}. Bar extremities on Fig.~\ref{fig:jitter-wind} indicate the 25 and 75 percentiles of the $\sigma_\eta$ values for a given wind speed.
As expected, strong winds produce large image jitter, but surprisingly,  $\sigma_\eta$ values remain lower than $5^{\prime\prime}$ for wind speeds lower than 30~$\mathrm{ms^{-1}}$. Even though the force exerted by the wind on the telescope is proportional to the wind speed, the mount and telescope structures move significantly only if the wind speed exceeds a certain value that seems to lie between 30 and 35~$\mathrm{ms^{-1}}$.

The telescope and mount stiffness, together with the guiding performance have proven to suffice for conducting observations in moderate to strong wind conditions, without the need of an auto guiding procedure like in the prototype LOLAS.

\section{ {\cn} Measurement examples}
\label{sec:profiles}

In this section we present a few examples of measured scintillation autocorrelations and corresponding turbulence profiles. The first results of LOLAS-2 were obtained in 2013 November at the OAN-SPM. The instrument was installed on a concrete pillar (see Fig.~\ref{fig:InstLOLAS}) to reduce vibrations.

\begin{figure}[h!]
\epsscale{.80}
\plotone{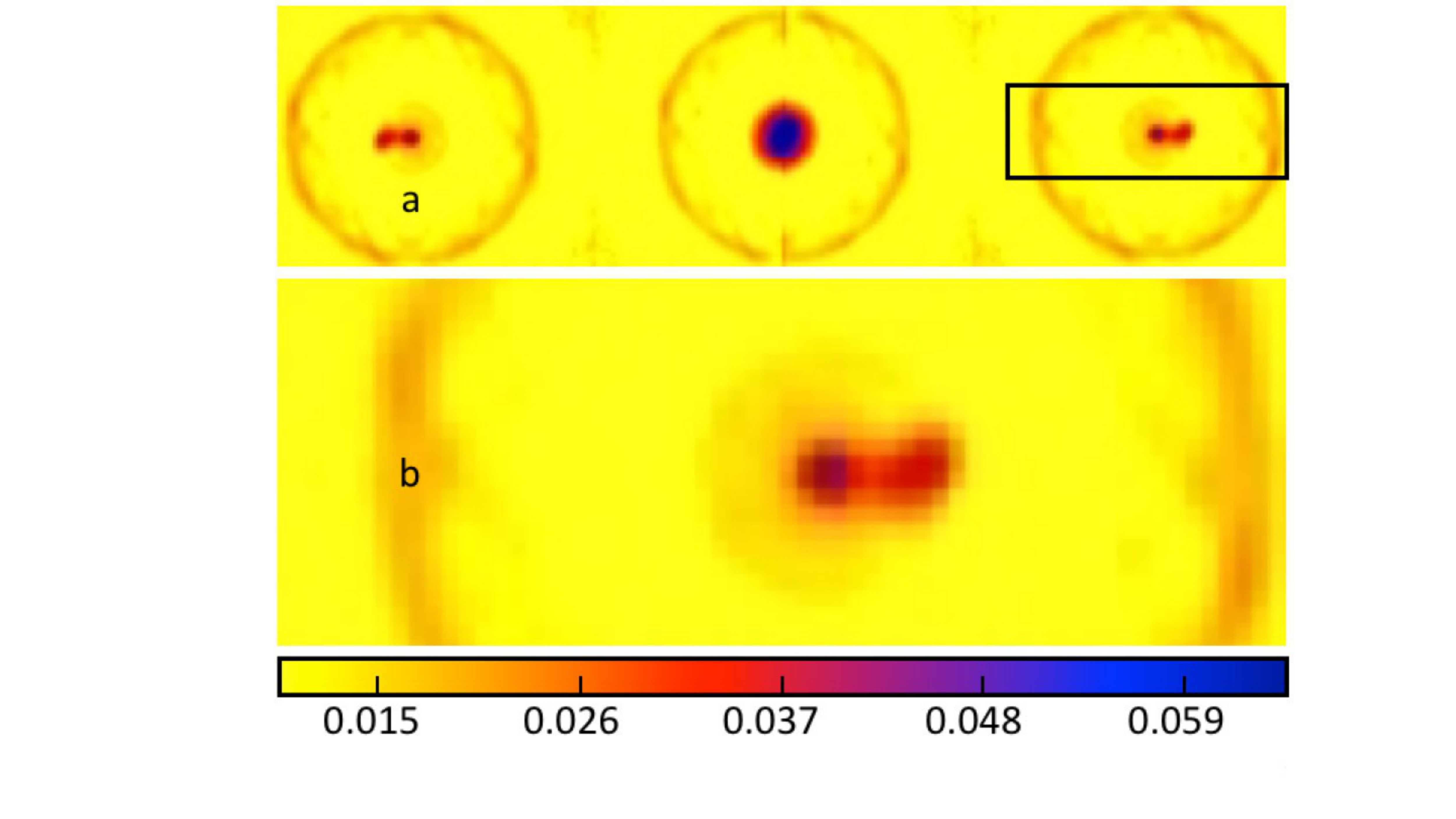}
\plotone{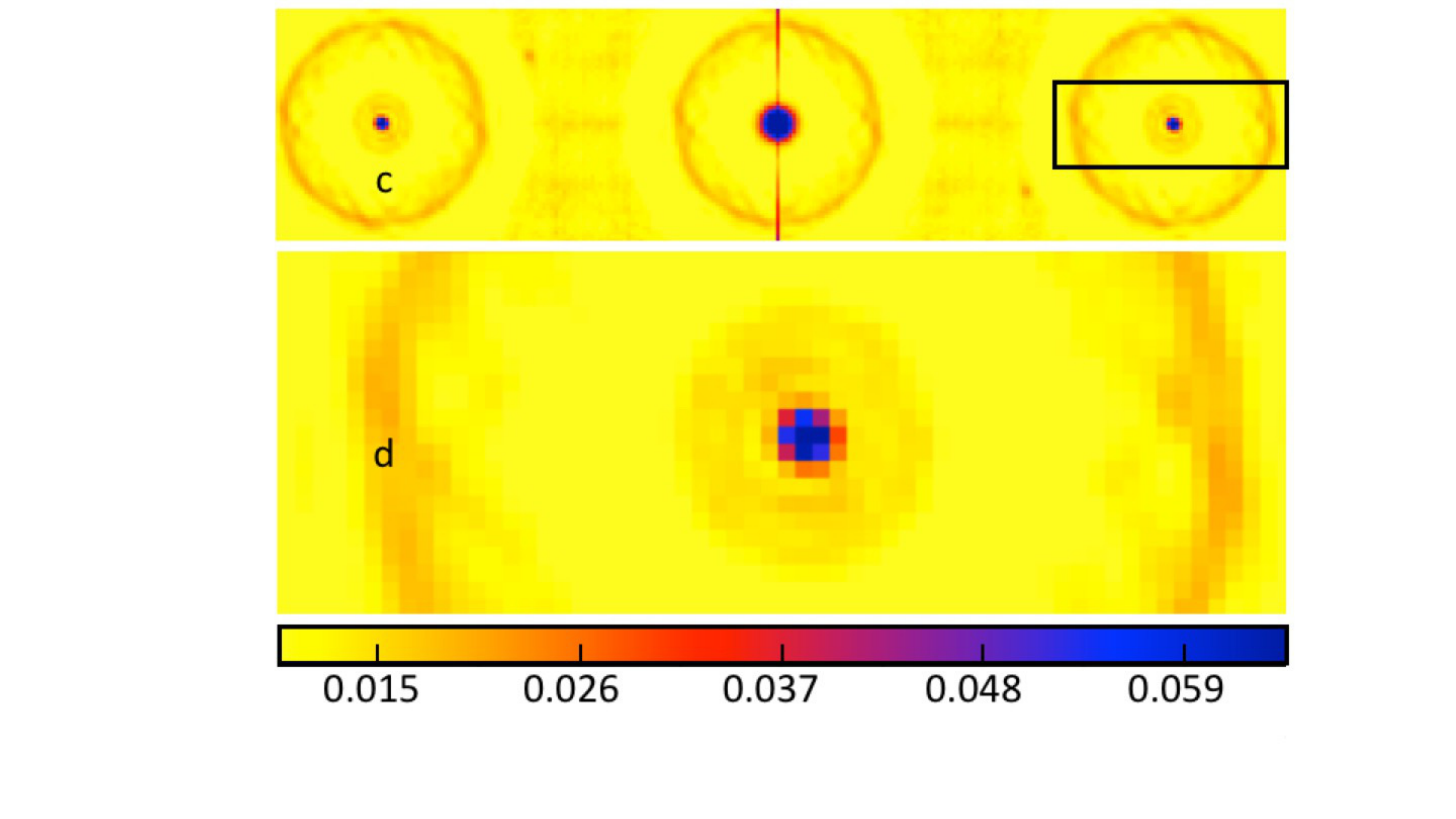}
\caption{Examples of measured autocovariance maps. Data were
obtained at the OAN-SPM on 2013 November 16 at 5:43 UT with the double
star 15 Tri (\textit{a} and \textit{b}) and on 2013 November 17 at 9:36 UT using
target  12 Cam ({\it c} and {\it d}).  Frames {\it b} and {\it d} show enlarged views of
the rectangles that appear in frames {\it a} and {\it c}.} \label{fig:ACs}
\end{figure}

Figure~\ref{fig:ACs} shows examples of autocovariance maps obtained from 30~000 scintillation images, during the nights 2013 November 16 and 17. In the central peaks of Figs.~\ref{fig:ACs}{\it a}  and ~\ref{fig:ACs}{\it c} , the contribution of all turbulent layers in the atmosphere are added up.
The correlated speckles produced by  turbulent layers above $h_{\mathrm{max}}$ are separated by a distance longer than the pupil diameter, thereby not giving rise to lateral peaks inside the autocovariance-map boundaries and avoiding the corresponding ${C_{\mathrm{N}}^2}$ estimate. Only the layers below $h_{\mathrm{max}}$ form visible lateral peaks from which ${C_{\mathrm{N}}^2}$ values are retrieved. A white rectangle frames the right-hand side lateral peaks in Figs.~\ref{fig:ACs}{\it a} and ~\ref{fig:ACs}{\it c}. Enlarged views are shown in Fig.~\ref{fig:ACs}{\it b} and ~\ref{fig:ACs}{\it d}. The most intense central peak inside each of those frames corresponds to the turbulence at ground level. Weaker peaks can clearly be seen in Fig.~\ref{fig:ACs}{\it a}, which correspond to turbulence above the ground. Fig.~\ref{fig:ACs}{\it d} only exhibits ground-level turbulence.

\begin{figure*}[h]
\epsscale{1}
\plotone{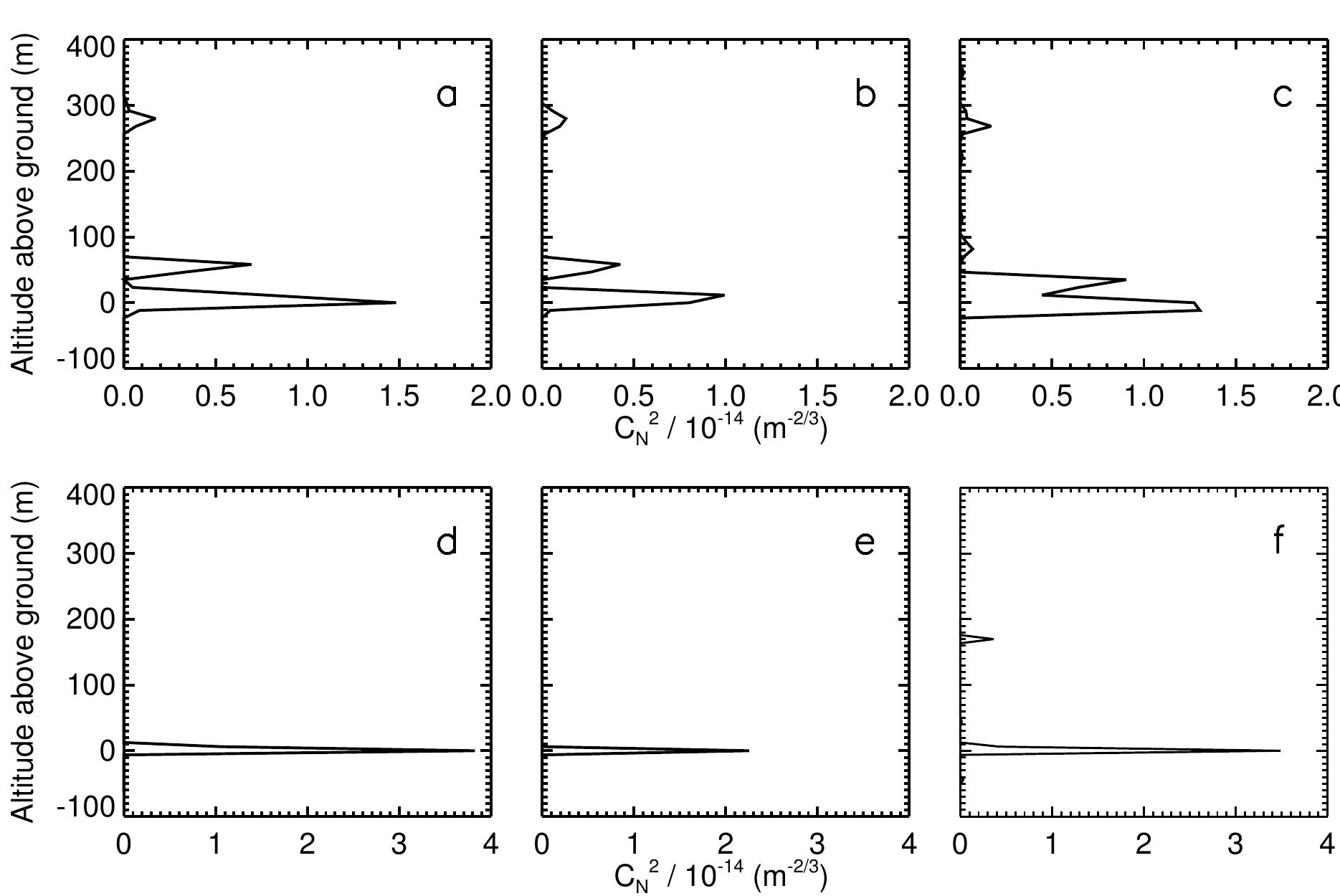}
\caption{Examples of {\cn} profiles obtained at the OAN-SPM on 2013 November 16 at 5:43 {\it a} , 5:47 ({\it b})  and 10:10 ({\it c})  and on 2013 November 17 at 9:36 ({\it d}) , 10:27 ({\it e})  and 11:27 ({\it f}) . Dates and times are in UT.}
\label{fig:profiles}
\end{figure*}

The  ${C_{\mathrm{N}}^2(h)}$ profiles are obtained using a modified CLEAN algorithm \citep{AAW+08} based on the one presented by \citet{PDA01}.
In Fig.~\ref{fig:profiles} we present a few profiles obtained on 2013 November 16 and 17 using autocovariance maps whose examples are  shown in
Fig~\ref{fig:ACs}. The purpose of Fig.~\ref{fig:profiles} is only to display the ability of the instrument in measuring ${C_{\mathrm{N}}^2}$ profiles and their characteristics. It is not intended for studying the optical turbulence at the site. Altitude resolution, maximum sensed altitude and noise level values are summarised in Table~\ref{tab:ParamsStars}. Note that the altitude resolution obtained when using 12 Cam as a target is the best ever achieved with Scidar-like techniques. The noise levels indicated in Table~\ref{tab:ParamsStars} are the {\cn} values that correspond to the $3\sigma$, i.e. three times the standard deviation on the autocovariance-map background \citep{AAW+08}. The profiles corresponding to November 16 show three turbulent layers: one at ground level, the second at 81 m and the weakest at  390 m. On November 17, almost all turbulence below 400 m was concentrated at ground level. Only a weak turbulent layer is detected at 170 m above the ground at 11:27 UT that night.  It is worth recalling that {\cn} values obtained with Scidar techniques in which the pupil images are not superimposed on the detector, like LOLAS, do not need to be corrected for the normalization error pointed out by \citet{AC09}.

\section{Conclusions}
\label{sec:conclusions}

The second generation of the Low Layer Scidar incorporates a simplified optical layout, a stiff and precise telescope mount and an open Ritchey-Chr\'etien telescope with excellent  focus stability. The analysis of the image positions showed that the guiding quality of the mount permits to avoid the autoguiding algorithm that was used in the prototype LOLAS. Similarly, the excellent focus stability of the telescope allows for observations without the autofocus algorithm employed in the former version of the instrument. Examples of measurements obtained with LOLAS-2 illustrate the capability of the instrument for very high altitude-resolution profiling of the optical turbulence. A statistical analysis of {\cnh} profiles obtained so far at the OAN-SPM will be presented in a forthcoming paper.

\acknowledgments
We are deeply grateful to the staff of the Observatorio Astron\'omico Nacional at San Pedro M\'artir (OAN-SPM) for their kind help in all the logistics for the observations. Wind  data used in the work was provided by the OAN-SPM wheather station (\url{http://tango.astrosen.unam.mx}). Financial support was provided by
DGAPA--UNAM through grants IN103913 and IN115013.

\clearpage

\clearpage

\clearpage

\clearpage

\clearpage

\clearpage

\clearpage

\clearpage

\epsscale{1}

\clearpage


\begin{deluxetable}{lcccc}
\tabletypesize{\scriptsize}
\tablecaption{Instrumental parameters for different values of $L$\label{tab:ComparisonL}}
\tablewidth{0pt}
\tablehead{
\colhead{Parameter} & \colhead{$L_1$=15 mm} & \colhead{$\bm{L_2=}${\bf 11 mm}} & \colhead{$L_3$=7.5 mm} & \colhead{$L_4$=5 mm}
}
\startdata
$h_\mathrm{gs}$[m]					& -886	& {\bf -1212}	& -1777	& -2668\\
$d$ [mm]\tablenotemark{a}			& 1.66	& {\bf 1.22}	& 0.83	& 0.55\\
$N_{d}$ [pixels]\tablenotemark{b}	& 52	& {\bf 38}		& 26	& 17\\
$\mathcal{L}_{D}/p$\tablenotemark{c}	& 2.58	& {\bf 1.80}	& 1.29	& 0.86\\
\enddata
\tablenotetext{a}{Diameter of the pupil image on the detector plane.}
\tablenotetext{b}{Number of binned pixels along a pupil image diameter (Fig. \ref{fig:ImageMean}).}
\tablenotetext{c}{Number of binned pixels along a speckle (for $h=0$).}
\end{deluxetable}

\clearpage

\begin{deluxetable}{lccccc}
\tabletypesize{\scriptsize}
\tablecaption{Targets\label{tab:UsedStars}}
\tablewidth{0pt}
\tablehead{
\colhead{Name} & \colhead{${\alpha_{2000}}$\tablenotemark{a}} & \colhead{${\delta_{2000}}$\tablenotemark{a}} & \colhead{${m_{1}}$\tablenotemark{b}} & \colhead{${m_{2}}$\tablenotemark{b}} & \colhead{$\rho$\tablenotemark{c} [$^{\prime \prime}$]}
}
\startdata
15 Tri & 2$^{\mathrm{h}}$:35 & 34$^{\circ}$:41 & 5.5 & 6.7 & 138.9\\
12 Cam & 5$^{\mathrm{h}}$:06 & 58$^{\circ}$:58 & 5.2 & 6.2 & 181.3\\
\enddata
\tablenotetext{a}{ Right ascension ($\alpha_{2000}$) and Declination ($\delta_{2000}$).}
\tablenotetext{b}{ Visible magnitude of each star.}
\tablenotetext{c}{ Angular separation.}
\end{deluxetable}

\clearpage

\begin{deluxetable}{lccccc}
\tabletypesize{\scriptsize}
\tablecaption{Observational parameters\label{tab:ParamsStars}}
\tablewidth{0pt}
\tablehead{
\colhead{Date} & \colhead{Target} & \colhead{$\Delta h$\tablenotemark{a}} & \colhead{${h_{\mathrm{max}}}$\tablenotemark{b}} & \colhead{$\tau$\tablenotemark{c}} & \colhead{N. L.\tablenotemark{d}}
}
\startdata
2013/11/16 & 15 Tri &  11.7 & 603 &  3   & 9.5$\times 10^{-16}$\\
2013/11/17 & 12 Cam   & 6.3 & 462  & 2   & 1.3$\times 10^{-15}$\\
\enddata
\tablenotetext{a}{ Altitude resolution [m].}
\tablenotetext{b}{ Maximum altitude [m].}
\tablenotetext{c}{ Exposure time [ms].}
\tablenotetext{d}{ Noise level [$\mathrm{m}^{-2/3}$].}
\end{deluxetable}

\clearpage

\end{document}